\documentclass[%
aps,
prl,
reprint,
superscriptaddress,
longbibliography,
amsmath,amssymb,
floatfix,
]{revtex4-2}
\usepackage[utf8]{inputenc}
\usepackage[T1]{fontenc}
\usepackage{graphicx}
\usepackage{bm}
\usepackage[colorlinks=True,linkcolor=blue,citecolor=blue,filecolor=blue,urlcolor=blue]{hyperref}
\usepackage{float}
\usepackage[normalem]{ulem}

\begin{document}

\title{Discovery of Higher-Order Topological Insulators using the \\ Spin Hall Conductivity as a Topology Signature}

\author{Marcio Costa}
\affiliation{Department of Physics, Fluminense Federal University, 24210-346, Niter\'oi, Rio de Janeiro, Brazil}
\affiliation{Brazilian Nanotechnology National Laboratory (LNNano), CNPEM, 13083-970 Campinas, Brazil}
\affiliation{Center for Natural and Human Sciences, Federal University of ABC, Santo Andr\'e, SP, Brazil}
\author{Gabriel R. Schleder}
\affiliation{Brazilian Nanotechnology National Laboratory (LNNano), CNPEM, 13083-970 Campinas, Brazil}
\affiliation{Center for Natural and Human Sciences, Federal University of ABC, Santo Andr\'e, SP, Brazil}
\author{Carlos Mera Acosta}
\affiliation{Center for Natural and Human Sciences, Federal University of ABC, Santo Andr\'e, SP, Brazil}
\author{Antonio C. M. Padilha}
\affiliation{Brazilian Nanotechnology National Laboratory (LNNano), CNPEM, 13083-970 Campinas, Brazil}
\author{Frank Cerasoli}
\affiliation{ Department of Physics and Department of Chemistry, University of North Texas, Denton TX, USA}
\author{M. Buongiorno Nardelli}
\affiliation{ Department of Physics and Department of Chemistry, University of North Texas, Denton TX, USA}
\author{Adalberto Fazzio}
\affiliation{Brazilian Nanotechnology National Laboratory (LNNano), CNPEM, 13083-970 Campinas, Brazil}
\affiliation{Center for Natural and Human Sciences, Federal University of ABC, Santo Andr\'e, SP, Brazil}

\begin{abstract}
The discovery and realization of topological insulators, a phase of matter which hosts metallic boundary states when the $d$-dimension insulating bulk is confined to ($d-1$)-dimensions, led to several potential applications. Recently, it was shown that protected topological states can manifest in ($d-2$)-dimensions, such as hinge and corner states for three- and two-dimensional systems, respectively. These nontrivial materials are named higher-order topological insulators (HOTIs). Here we show a connection between spin Hall effect and HOTIs using a combination of {\it ab initio} calculations and tight-binding modeling. The model demonstrates how a non-zero bulk midgap spin Hall conductivity (SHC) emerges within the HOTI phase. Following this, we performed high-throughput density functional theory calculations to find unknown HOTIs, using the SHC as a criterion. We calculated the SHC of 693 insulators resulting in seven stable two-dimensional HOTIs. Our work guides novel experimental and theoretical advances towards higher-order topological insulators realization and applications.
\end{abstract}

\maketitle

\textit{Introduction$-$}
The discovery of symmetry protected topological phases of matter opened a new field in condensed matter physics. 
When a $d$-dimensional topological material is confined to $(d-1)$-dimension, symmetry protected gapless states, which are robust against perturbations, manifest at its boundaries~\cite{QSH,RevModPhys.82.3045,RevModPhys.83.1057}, named first-order topological insulator (TI).  This fundamental characteristic of TIs is the so-called bulk-boundary correspondence principle~\cite{JPSJ.81.114602}, which is mediated through a variety of symmetries, including time-reversal (TR)~\cite{QSH}, mirror planes~\cite{TCI}, non-symmorphic~\cite{pssr.201206451}, and particle-hole~\cite{Particlehole}.
Recently, a novel category of nontrivial material that apparently violates the bulk-boundary correspondence was proposed, which is referred to as higher-order topological insulators (HOTIs)~\cite{Benalcazar2017,Schindlereaat0346}. Specifically, HOTIs exhibit \textit{gapped} states in the $d-$(bulk) and $(d-1)$ $-$dimensional boundaries, although exhibiting \textit{gapless} protected states at the $(d-2)$ boundaries. For instance, the three-dimensional (3D) bismuth crystal is a HOTI protected by the three-fold rotational ($R_{3}$), TR, and inversion (IS) symmetries. 
For a one-dimensional hexagonal bismuth nanowire which preserves all three symmetries, hinge gapless protected states are observed~\cite{Schindler2018}.
Despite the abundance of first-order TIs in nature~\cite{Vergniory_HT_TI,Tang_HT_TI,Zhang_HT_TI}, HOTIs are seemingly less common\cite{Ezawa2018prl,graphyne,graphene_bilayer,Graphdiyne}, indicating a need to develop novel strategies or principles capable of discovering and design of such systems. 
 
Angle-resolved photoemission spectroscopy (ARPES) is a frequent approach to identify first-order TIs, where the boundary states are directly accessible in $(d-1)$-dimension. Another experimental method is measuring the spin currents generated by the spin Hall effect (SHE)~\cite{spinCurrent1,spinCurrent2}. 
Since spin Hall currents and spin Hall conductivity (SHC) are intimately related~\cite{spinCurrent}, this strategy is also useful from the theoretical point of view, where the latter can be calculated via the Kubo formula or via the real space local spin Hall conductivity~\cite{LSHC}.
In general, undoped insulators show zero midgap SHC~\cite{KM1}.
Murakami {\it et al.} proposed a strategy to obtain finite SHC in narrow-gap insulators such as HgS and PbSe~\cite{SHC-Murakami}, which later were identified as first-order TIs \cite{Vergniory_HT_TI}, suggesting for the first time a connection between the SHE and nontrivial topology. 
The SHC in first-order TIs is given by $\sigma^{z}_{xy}=C_{s}e^{2}/h$, where $C_{s}$ is the so-called spin Chern number, defined as the first Chern number difference for each spin channel ($C_{s}= C_{\uparrow}-C_{\downarrow}$). 
Ideally, these materials possess a perfect quantized SHC. Nevertheless, in general, small deviations occur due to the absence of conservation of the spin angular momentum $z$-component ($S_z$), owing to effects such as crystal field, hybridization, symmetry breaking terms \cite{QSH}, or disorder~\cite{KM1}. Conversely, the relation between the SHC and HOTI materials has not been explored so far.
\begin{figure*}[ht!]
\includegraphics[width=\textwidth]{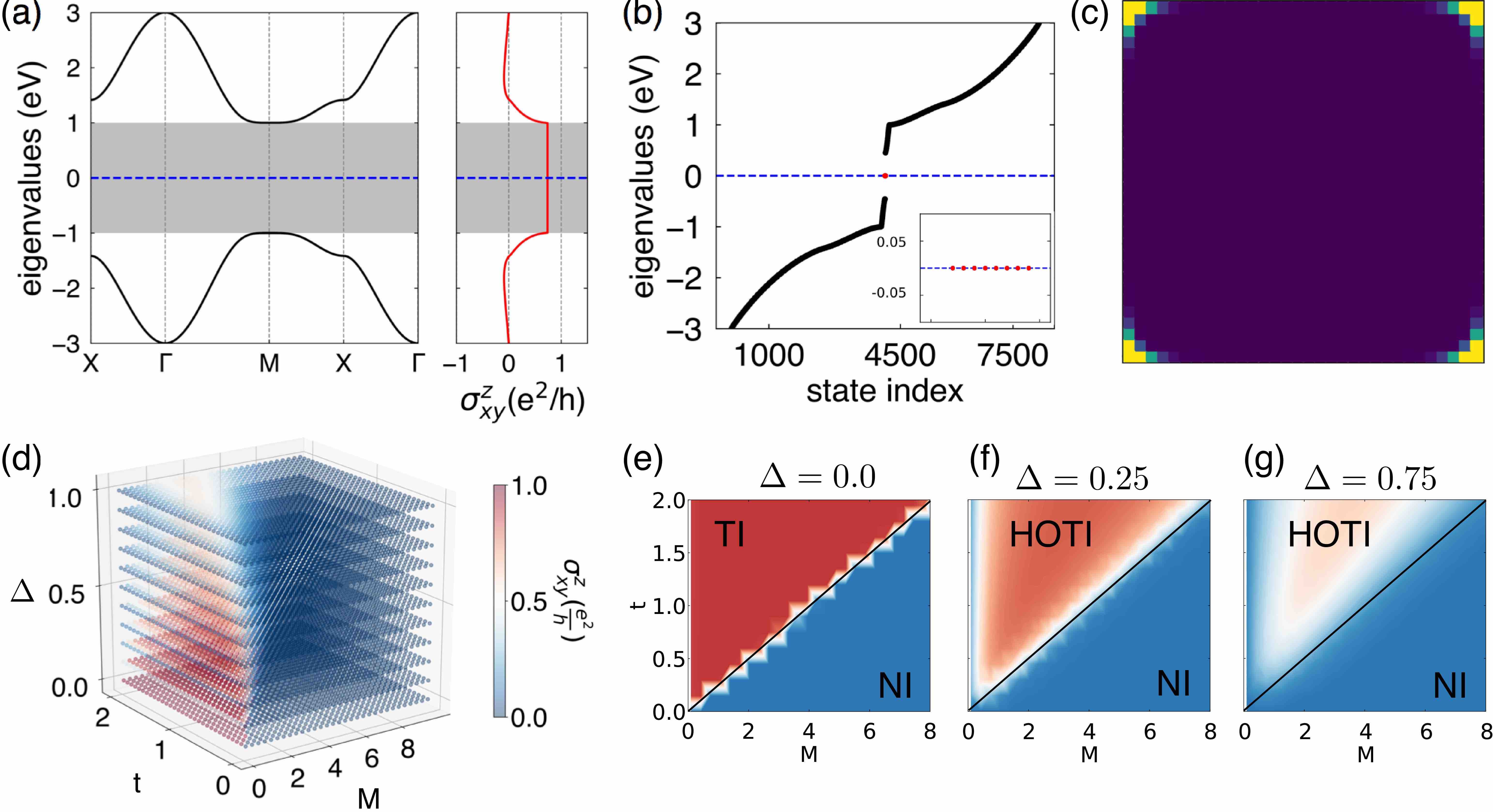}
\caption{\label{model-SHC} { Two-dimensional higher-order topological insulator tight-binding model.} ({a}) Bulk band-structure and spin Hall conductivity ($\sigma_{xy}^{z}$) as a function of the Fermi energy. ({b}) Eigenvalues for a square flake geometry (open boundary conditions) with 32$\times$32 sites. The red circles represents the degenerate corner states, with the inset showing a narrow energy window to emphasize their degeneracy. ({c}) Corner states real-space projection. The model parameters of equation \ref{TB-SQ} are $M = 2t = 2\xi = 1$ and $\Delta = 0.25$. ({d}) Spin Hall conductivity as a function of the model parameters ($0 \leq t \leq 2$), ($0 \leq\Delta\leq 1$) and ($0 < M\leq 10$). ({e}), ({f}), and ({g}) Spin Hall conductivity for fixed values of $\Delta=$ 0, 0.25, and 0.75, respectively.} 
\end{figure*}

Here we demonstrate that the SHC is a signature of HOTI phases. For this purpose, we first employ a tight-binding model to elucidate the conditions leading to a HOTI phase and its connection with a large bulk-SHC inside the topological gap. With the connection between finite midgap SHC and HOTI phases established, we perform a systematic search for novel HOTIs. We calculate the SHC of 693 insulators reported in the 2D materials database C2DB~\cite{c2db}, followed by a screening procedure consisting of the selection of only thermodynamically stable non-magnetic insulating materials. 
Among the possible candidates, we discover seven stable two-dimensional HOTIs: BiSe, PbF, PbBr, PbCl, HgTe and BiTe in both GaSe and CH prototype structures.  
Real-space projections of midgap states obtained from DFT calculations show symmetry-protected localization of topological states in nanoflakes of these systems. We then compute the topological invariant for the representative candidate BiSe, validating our predictions. 
Our work advances the understanding of HOTI phenomena and demonstrates an important correspondence with the SHE.

\textit{SHC as a signature of HOTIs$-$}
Initially, we demonstrate how a HOTI shows a non-zero midgap SHC. We start from an eight-band tight-binding Hamiltonian, proposed by Schindler {\it et al.} in ref.~\cite{Schindler2018}. The model is adapted for a two-dimensional material and is given by
\begin{align}
\label{TB-SQ}
H  = \sum_{i,e=x,y} & c^{\dagger}_{i}(t\tau_{z}\otimes\rho_{0}\otimes\sigma_{0}+i\xi \tau_{x}\otimes\rho_{0}\otimes\sigma_{e} )c_{i+e} + {\it h.c.} \nonumber \\
 + \sum_{i,e=x,y} & c^{\dagger}_{i}(\Delta d_{e} \tau_{y}\otimes\rho_{y}\otimes\sigma_{0} )c_{i+e} + {\it h.c.} \nonumber \\
+ \sum_{i}\ & c^{\dagger}_{i}(M\tau_{z}\otimes\rho_{0}\otimes\sigma_{0})c_{i},
\end{align}
where we consider two full spin orbitals $\mu$ and $\nu$ on each site. The site index is represented by $i$ and $c^{\dagger}_{i}$ ($c_{i}$) is the creation (annihilation) fermionic operator. The Pauli matrices $\rho$, $\tau$, and $\sigma$ stands for the orbital $\mu$, $\nu$ and spin, respectively. The term $d_{e} = \pm1$ for $e = x,y$, while $M$, $t$, $\xi$ and $\Delta$ are the mass, hopping, spin-orbit coupling (SOC) strength and mixing term, respectively. This model realizes a 2D HOTI that preserves TR-symmetry ($\mathcal{T} = -is_{y}K$) and four-fold rotational ($R_{4}$) symmetry ($C_{4}=\tau_{z}e^{-i\pi s_{z}/4}$). 
Setting the parameters $M=2t=2\xi=1$ and $\Delta=0.25$, the bulk (2D) band structure with a 2 eV bandgap is obtained as seen in Fig.~\ref{model-SHC}({a}). In a torus geometry, a gapped band structure is obtained (see Supplementary Information Fig.~S1), typical of HOTIs since its nontrivial topology is manifested in ($d-2$). Fig.~\ref{model-SHC}({b}) shows the $0D$ energy spectrum, {\it i.e.}, square nanoflake geometry (open boundary conditions). The inset highlights the zero energy modes (red circles) which are localized at the flake corners ($d-2$ boundary states), as shown in Fig.~\ref{model-SHC}({c}). In Fig.~\ref{model-SHC}({a}) side plot, we show the calculated HOTI model SHC $\sigma^z_{xy}\approx 0.75\, e^{2}/h$, showing that SOC introduces a non-zero SHC into HOTI phases. 
To deepen our analysis, we calculate the SHC landscape in a given range of model parameters, $0 \leq t \leq 2$, $0 \leq\Delta\leq 1$, and $0 < M\leq 10$, keeping the SOC strength fixed at $\xi=0.5$. The resulting map in Fig.~\ref{model-SHC}({d}) shows a clear inverse relation between the SHC and $\Delta$. To clarify their relationship we analyze cross sections of the SHC landscape in Figs.~\ref{model-SHC}({e}-{g}). 
For $\Delta=0$ (Fig.~\ref{model-SHC}({g})), we obtain a TI with a quantized SHC in the blue region. The phase boundary between the TI and NI region depends on the $|M|/t$ ratio and is depicted as a black line. For $\Delta=0.25$ the HOTI phase is obtained, as can be seen in the red region, where the SHC is no longer quantized, which is evidenced by its pale color. Such reduction in SHC is more pronounced as $\Delta$ increases, according to Fig.~\ref{model-SHC}({g}), where $\Delta=0.75$. As previously stated, the $\Delta$ term is responsible for driving a TI into the HOTI phase. This analysis of the model Hamiltonian evidences the gradual suppression of the SHC as the system goes further into the HOTI phase, $i.e.$, when $\Delta$ increases.

\textit{First principle calculations$-$}
We performed a screening over the 2D materials database C2DB~\cite{c2db}. The C2DB is constructed based on prototypes of known 2D materials, {\it i.e.}, Graphene, MoS$_{2}$, BN, and others. With these prototypes, a combinatorial approach created around 4000 2D compounds for which the properties were obtained via density functional theory (DFT) calculations. The authors established criteria to categorize the materials as low, medium, and high stability. Thermodynamic stability is evaluated by computing each candidate's convex Hull. The dynamic stability is based on the phonon spectra of experimentally synthesized 2D materials, see ref.~\cite{c2db} for details. 
We performed a screening over the C2DB database to select materials with appropriate characteristics. Initially, we selected thermodynamic and dynamic stable materials. After this initial screening, we narrowed our search over non-magnetic insulators, resulting in 693 2D materials.

Following the initial screening we performed fully relativistic density functional theory~\cite{DFT1,DFT2,review} calculations using the \textsc{Quantum Espresso} package~\cite{QE-2017}. The exchange-correlation (XC) is described within the generalized gradient approximation (GGA) via the Perdew-Burke-Ernzerhof (PBE) functional~\cite{pbe}. Ionic potentials were described by projector augmented-wave (PAW)~\cite{PAW} pseudopotentials available in the pslibrary database~\cite{pslibrary}. Both the GGA exchange-correlation functional and the PAW ionic pseudopotentials are the same as used in the C2DB database construction. The wavefunctions and charge density cutoff energy were 40\% larger than the recommended. The reciprocal space sampling was performed with a $K$-point density of 6.0/\AA$^{-1}$ for structural optimization and 12.0/\AA$^{-1}$ for self-consistent calculations. The lattice parameters reported in the database were used, and a full structural optimization was performed until Hellman--Feynman forces were smaller than 0.01 eV/\AA. 
We calculated the spin Hall conductivity (SHC) of all selected 2D materials and validated our results with the C2DB topological classification reported in ~\cite{c2db-TI}. 
Then, we further selected all materials with large SHC and not classified as topological insulators or topological crystalline insulators.
Finally, for the predicted HOTIs we calculated the electronic structure of triangular nanoflakes, corresponding to open boundary conditions, using the \textsc{vasp} software code~\cite{Kresse96}. To avoid spurious interactions, 15\AA{} of vacuum is used in all calculations.

The SHC was calculated using local effective Hamiltonians constructed via the pseudo-atomic orbital (PAO) projection method~\cite{PAO1,PAO3}, implemented in the \textsc{paoflow} code~\cite{PAO5}. The method consists in projecting the plane wave Kohn-Sham states, composed of several thousand of basis functions, into a compact subspace spanned by PAO orbitals.
The PAO Hamiltonian allows the efficient and accurate computation of the  SHC~\cite{jagoda}, topological properties~\cite{Costa2019,TI-NI}, dynamical properties~\cite{dynamics} and others. 
A comparison between the DFT and PAO electronic structures is presented in the Supplementary Information Fig.~S6.
The SHC was calculated via linear response theory using the Kubo formula,
\begin{eqnarray} 
\label{SHC}
\sigma^s_{ij}=\frac{e^2}{\hbar}\sum_{\vec{k}}\sum_{n}f(\vec{k})\Omega^{s}_{n,ij}(\vec{k}),
\end{eqnarray}
where $s$ is the quantization axis (spin polarization), $j$ is the applied electrical field orientation and $i$ gives the spin current direction ($\sigma_{ij}^s$). The Fermi-Dirac distribution is represented by $f(\vec{k})$ 
\begin{figure}[hb!]
\includegraphics[width=\linewidth]{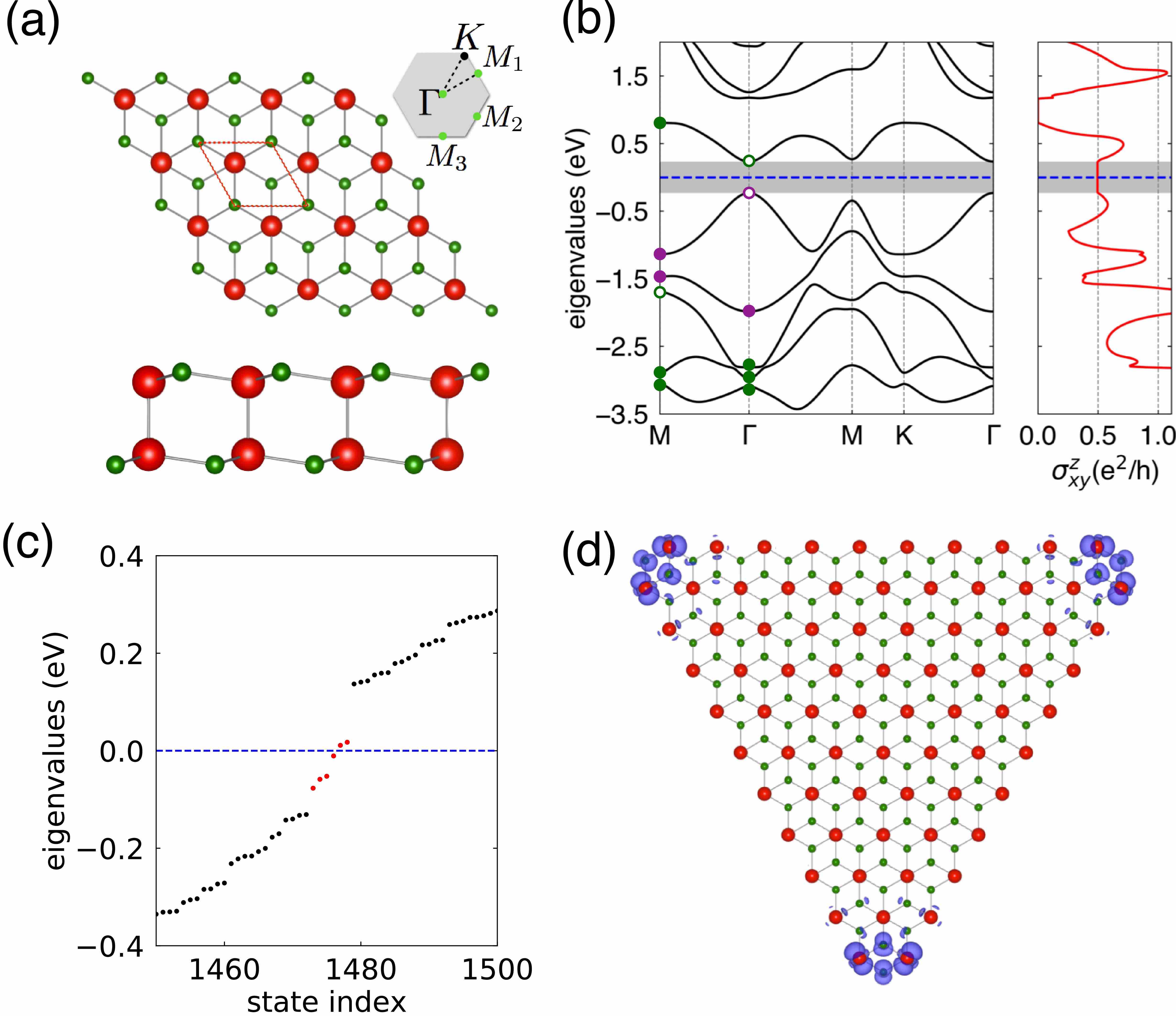}
\caption{\label{bulk-SHC} {Two-dimensional HOTI material BiSe (in the GaSe prototype).} ({a}) 2D bulk crystal structure top and side views. The inset shows the Brillouin zone high-symmetry points. ({b}) Bulk band-structure along the high symmetry lines and the spin Hall conductivity ($\sigma^{z}_{xy}$) as a function of the energy. The dashed blue line represents the Fermi level. The gray shaded area encompass the bulk band gap. Green and purple dots represent states with $R_{3}$ eigenvalues $\gamma_{R_{3}}=-\pi$ or $\pm\pi/3$, respectively. Filled and unfilled dots represent $+$ and $-$ parity eigenvalues ($\xi_{i}=\pm$ 1) for each occupied band at the TRIM $K$-points.
({c}) Energy eigenvalues of a triangular flake with edges of approximately 50~\AA. The localized corner modes are highlighted in red. ({d}) Flake structure with corner states charge density highlighted in blue.}
\end{figure}
and $\Omega^{s}_{n,ij}(\vec{k})$ is the spin Berry curvature given by
\begin{eqnarray} 
\label{Berry}
\Omega^{s}_{n,ij}(\vec{k})=\sum_{m\neq n}\frac{2\mathrm{Im}\langle\psi_{n,\vec{k}}|j_{i}^{s}|\psi_{m,\vec{k}}\rangle\langle\psi_{m,\vec{k}}|v_{j}|\psi_{n,\vec{k}}\rangle}{(E_{n} - E_{m})^2} ,
\end{eqnarray}
where $m$ and $n$ are band indices, $\psi$ and $E$ are the PAO Hamiltonian wavefunction and eigenvalues, respectively. The spin current operator ($j_{i}^{s}$) is defined as the anticommutator of the Pauling matrix ($\sigma_{s}$) and velocity operators ($\{\sigma_{s},v_{i}\}$). 
For the SHC the Brillouin zone sampling was increased to 72.0/\AA$^{-1}$.
The use of non-degenerate perturbation theory in the Kubo formula can lead to unphysical midgap SHC in trivial insulators~\cite{GaAs}. 
To circumvent this issue \textsc{paoflow} uses degenerate perturbation theory, see ref.~\cite{jagoda} for details.
In the case of spin rotation symmetry breaking, the spin quantization axis is not well defined, leading to a reduction of the midgap SHC in TIs~\cite{QSH,Matusalem}. 
Nevertheless, it does not induce unphysical midgap SHC in trivial insulators. In our calculations, the only nonzero midgap SHCs are due to TIs or HOTIs. The Hamiltonian term responsible for the higher-order topology in inversion symmetry compounds also induces the fractional SHC. This suggests that the fractional SHC is an indicator of higher-order topology in compounds that are not topological insulators. Our approach is applicable to HOTIs having significant SOC, see Fig. S5. 
Analog to topological crystalline insulators, since SOC is not explicitly required for HOTIs, the recently predicted graphdiyne ~\cite{Graphdiyne} and graphyne~\cite{graphyne}  having negligible SOC would not be identified.

\textit{BiSe electronic structure$-$}
Next, we focus on the electronic structure of a representative candidate, BiSe on the GaSe prototype (BiSe-GaSe). In Fig. \ref{bulk-SHC}({a}) the BiSe-GaSe crystal structure is displayed. It features inversion, TR, and three-fold rotational symmetry. The inset presents its Brillouin zone and  TR invariant momenta (TRIM) points. The fully relativistic band structure, Fig. \ref{bulk-SHC}({b}), reveals a direct bandgap of 0.46 eV at the $\Gamma$ point. Its SHC clearly exhibits a constant value of $\sigma^z_{xy}\approx 0.5 \,e^{2}/h$ in the bandgap region. 
Additionally, a distinguishing characteristic of a HOTI is the appearance of ($d-2$) (in this case, corner) states inside the bulk bandgap whenever the ($d-2$) geometry preserves the underlying symmetries. As such, since BiSe is protected by $C_3$, we constructed a triangular nanoflake with edges of approximately 50 \AA{}, shown in Fig. \ref{bulk-SHC}({d}). 
Along with the nanoflake geometry, the real-space wave function projection of the six eigenstates closest to Fermi level is shown, evidencing its localization at the corners. The nanoflake eigenvalues are displayed in Fig. \ref{bulk-SHC}({c}), where the corner states are highlighted in red. 
The expected degeneracy is lifted owing to the interaction between corner states, as previously reported for other system~\cite{graphyne}. A detailed discussion of interaction and localization of corner states is presented in the Supplementary Information (Figs. S2, S3, S4, and S5).
Indeed, using the local SHC, Rauch and Töpler have shown that bulk and boundary states give different contributions to the SHC \cite{LSHC}.


\textit{Topological invariant$-$}
To definitively demonstrate BiSe higher-order topological properties we introduce an adaptation for two dimensions of the invariant defined in ref. \cite{Schindler2018} for 3D HOTIs. The symmetry protected topological phases are driven by band inversions at $K$-points that preserve the symmetries protecting the bulk topology~\cite{RevModPhys.88.021004}. Such band inversions are characterized by a topological invariant~\cite{RevModPhys.88.035005,PhysRevB.76.045302}.
For instance, in inversion symmetry (IS) insulators protected by the TR-symmetry, band inversions are characterized by the parity eigenvalues product ($\xi_{i}=\pm$ 1) for all occupied bands at all TRIM points (the $Z_{2}$ topological invariant)~\cite{KM1}.
Similarly, in IS-HOTIs protected by $R_{3}$ and TR symmetries, band inversions at high-symmetry $K$-points preserving those symmetries can be divided into two groups of states according to the $R_{3}$ eigenvalue ($\gamma_{R_{3}}=-\pi$ or $\pm\pi/3$), see Fig. \ref{bulk-SHC}({b}). For each band inversion, there is an invariant given by the products of the IS eigenvalues for all occupied bands, {\it i.e.}, $\nu^{\gamma_{R_{3}}}_{K}=\prod_{i\in \text{occ}} \xi^{\gamma_{R_{3}}}_{i}$~\cite{RevModPhys.88.035005,PhysRevB.76.045302}. The band inversion for each $R_{3}$ eigenvalue is finally given by the product of the invariants for individual $K$-points, {\it i.e.}, $\nu^{\gamma_{R_{3}}}=\prod_{K} \nu^{\gamma_{R_{3}}}_{K}$, leading to either a trivial ($\nu^{\gamma_{R_{3}}}=1$) or topological ($\nu^{\gamma_{R_{3}}}=-1$) band structure. 
Subsequently, the invariant characterizing the topological states of the whole system is given by the product $\nu=\nu^{-\pi}\nu^{\pm\pi/3}$. 
This defines three possible topological states: \textit{(i)} $\nu^{-\pi}=\nu^{\pm\pi/3}=1$ for trivial insulators, \textit{(ii)} $\nu^{-\pi}=\nu^{\pm\pi/3}=-1$ for 2D HOTIs, and \textit{(iii)} $\nu^{-\pi}\neq\nu^{\pm\pi/3}$ for $Z_{2}=1$ topological insulators. For this reason, materials screening based on the invariant $\nu$ identifies 2D HOTIs ($\nu=1$) as trivial insulators~\cite{c2db-TI,marrazzo2019abundance}.
The Brillouin zone of 2D materials host four TR-symmetry invariant $K$-points. 
For instance, in the BiSe Brillouin zone, the high symmetries $K$-points $\Gamma$ and $M_{1,2,3}$ preserve TR- and $R_{3}$ symmetries (See Fig. \ref{bulk-SHC}({a})), meaning that bulk band inversion at these $K$-points can simultaneously be protected by both symmetries.
The topological indexes at the $M_1$, $M_2$, and $M_3$ points, which are mapped into each other by the $R_{3}$ rotation symmetry, only affect the global index for states identified as $-\pi$. This allows us to write the invariants as $\nu^{-\pi}=\nu_{\Gamma}^{-\pi}\nu_{\text{M}_{1}}$ and $\nu^{\pm\pi/3}=\nu_{\Gamma}^{\pm\pi/3}$. 
As represented in Fig. \ref{bulk-SHC}({b}), the IS eigenvalues 
leads to $\nu^{-\pi}=\nu^{\pm\pi/3}=-1$, indicating a HOTI phase with corner states charge density protected by the $R_{3}$ and TR-symmetries (Fig.  \ref{bulk-SHC}({d})). 

\begin{figure}[ht!]
\includegraphics[width=\linewidth]{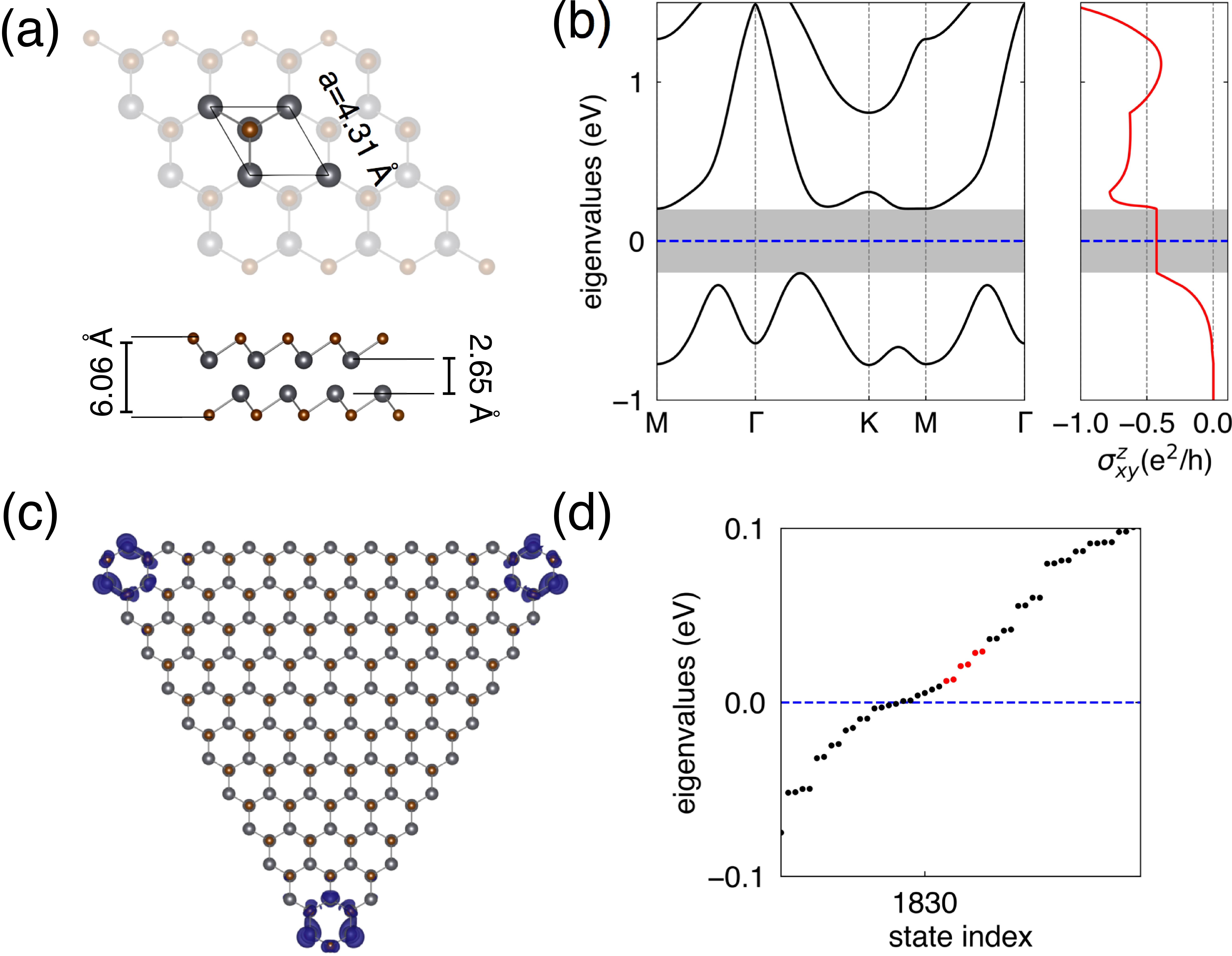}
\caption{\label{bulk-SHC-4materials} PbBr in the CH prototype: ({a}) crystalline structure, ({b}) bulk band structure and spin Hall conductivity, ({c}) flake structure with corner states charge density, and ({d}) corresponding 0D eigenvalues.}
\end{figure}

\textit{Predicted 2D HOTIs$-$}
Finally, the detailed analysis is repeated for the other discovered HOTIs. The predicted materials present the structural prototypes GaSe and CH, which correspond to AB stacking centrosymmetric bilayers preserving the TR- and $C_{3}$ symmetries. The compounds are formed by elements with high intrinsic SOC, such as Bi and Pb. 
This results in \textit{i}) small bulk bandgaps, i.e., 260, 406, 58, 398, 57 and 94 meV for BiTe, PbBr, BiTe-CH, PbCl, PbF and HgTe respectively; and \textit{ii}) large bulk SHC. In Fig.~\ref{bulk-SHC-4materials} we show the crystal structure, bulk band structure and SHC, flake structure with corner states charge density, and corner states eigenvalues for the PbBr in the CH prototype. As expected, we obtain a large midgap SHC of $\approx -0.5 e^{2}/h$ and midgap states localized at the triangular flake corners. The other materials are shown in the Supplementary Information, see Figs.~S7,~S8, ~S9, ~S10 and ~S11.
Despite not being quantum spin Hall insulators, the midgap SHC is a constant fraction of the quantum conductivity $e^{2}/h$. These constant values of SHC are different for compounds formed by Bi and Pb atoms. Nevertheless, the magnitude of the SHC is strongly dependent on the structure. For instance, the same BiTe composition in the structural prototypes GaSe and CH have SHC of 0.75 and 0.3 $e^{2}/h$, respectively.

\textit{Conclusion$-$}
We have shown the relation between the spin Hall effect and the higher-order topological insulator (HOTI) phase. Our tight-binding model predicts a non-zero bulk midgap spin Hall conductivity (SHC). This finding allowed us to search for novel HOTIs. We combined density functional theory calculations with local effective Hamiltonians to calculate the SHC of 693 insulators. We found seven stable two-dimensional HOTIs candidates: BiSe, BiTe (in two different structures), PbF, PbBr, PbCl and HgTe. All discovered HOTIs display metallic states in the bulk bandgap for ($d-2$)$-$dimensional (0D) structures preserving the $R_{3}$ symmetry, being spatially localized at the corners. 
This confirms the existence of the HOTI states protected by the $R_{3}$ symmetry and evidence the relationship between the $d-$dimensional bulk SHC and the topologically protected states in ($d-2$) dimensions. 
While we establish numerically the relation between the SHC and the higher-order topological phase, the development of analytical expressions for the SHC in HOTI models and implementations of HOTI invariants suitable for high-throughput calculations are both desirable.
Our work advances the understanding of higher-order topological phases showing for the first time its connection with the spin Hall effect.

\textit{Note$-$} After submission of this work, we became aware of a related study by Peterson \textit{et al.}, performing the identification of higher-order metamaterials systems by the fractional corner anomaly \cite{Peterson2020}.

\textit{Acknowledgments$-$} MC, GRS, CMA, ACMP, and AF acknowledges financial support from the Funda\c{c}\~ao de Amparo \`a Pesquisa do Estado de S\~ao Paulo (FAPESP), project numbers 16/14011-2, 17/18139-6, 18/11856-7, 18/05565-0, 17/02317-2. MC and MBN acknowledge Dr. Jagoda S\l awi\'{n}ska for useful discussions. The authors acknowledge the National Laboratory for Scientific Computing (LNCC/MCTI, Brazil) for providing HPC resources of the SDumont supercomputer and SAMPA/USP for the for providing HPC resources of the Josephson computer.
MBN also acknowledges the High Performance Computing Center at the University of North Texas and the Texas Advanced Computing Center at the University of Texas, Austin, for computational resources.

\normalbaselines
\bibliography{main}
\end{document}


\title{\Large{Supplementary Information} \\ \vspace{0.5cm}
\large{\textit{
Discovery of Higher-Order Topological Insulators using the \\ Spin Hall Conductivity as a Topology Signature
}}
}

\author{Marcio Costa}
\affiliation{Department of Physics, Fluminense Federal University, 24210-346, Niter\'oi, Rio de Janeiro, Brazil}
\affiliation{Brazilian Nanotechnology National Laboratory (LNNano), CNPEM, 13083-970 Campinas, Brazil}
\affiliation{Federal University of ABC, Santo Andr\'e, SP, Brazil}
\author{Gabriel R. Schleder}
\affiliation{Brazilian Nanotechnology National Laboratory (LNNano), CNPEM, 13083-970 Campinas, Brazil}
\affiliation{Federal University of ABC, Santo Andr\'e, SP, Brazil}
\author{C. Mera Acosta}
\affiliation{Federal University of ABC, Santo Andr\'e, SP, Brazil}
\author{Antonio C. M. Padilha}
\affiliation{Brazilian Nanotechnology National Laboratory (LNNano), CNPEM, 13083-970 Campinas, Brazil}
\author{Frank Cerasoli}
\affiliation{ Department of Physics and Department of Chemistry, University of North Texas, Denton TX, USA}
\author{M. Buongiorno Nardelli}
\affiliation{ Department of Physics and Department of Chemistry, University of North Texas, Denton TX, USA}
\author{Adalberto Fazzio}
\affiliation{Brazilian Nanotechnology National Laboratory (LNNano), CNPEM, 13083-970 Campinas, Brazil}
\affiliation{Federal University of ABC, Santo Andr\'e, SP, Brazil}

\maketitle
\clearpage
\subsection{Tight$-$Binding Model : TI $\times$ HOTI}
\begin{figure*}[ht!]
\includegraphics[width=\textwidth]{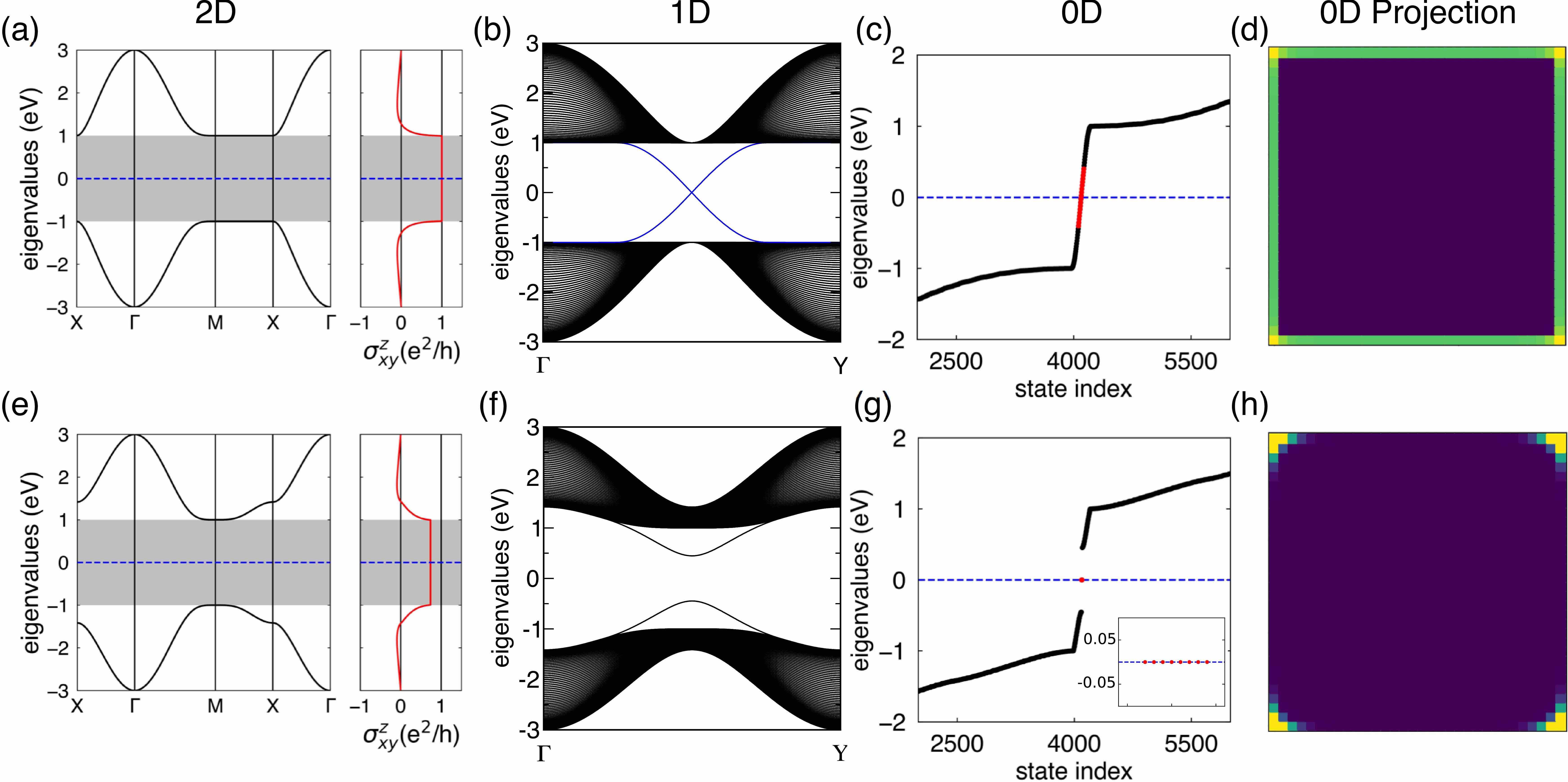}
\caption{\label{HOTIxTI} 2D tight binding model: TI $\times$ HOTI. Comparison between the TI ($\Delta=0.0$) (a-d) and HOTI ($\Delta=0.25$) (e-h) phases. Tight-binding model parameters are $M=2t=2\xi=1$. (a,e) 2D Bulk band structure along with the spin Hall conductivity, (b,f) 1D nanoribbon band structure, (c,g) 0D nanoflake eigenvalues, and (d,h) midgap states wave function real space projection (edge states for TI and corner states for HOTI).}
\end{figure*}

The tight-binding model described in the main text represents a 2D square lattice with an onsite mass term ($M$), nearest-neighbor hopping ($t$), intrinsic spin-orbit coupling ($\xi$) and a mixing term ($\Delta$). 
Exploring the model parameters one can go from a normal insulator (NI), $|M|/t > 3$ and $|M|/t < 1$, to a topological insulator (TI), $1 < |M|/t < 3$. 
Within the TI phase, for any nonzero value of the mixing term ($\Delta > 0$) the HOTI phase is obtained. 
We used the following model parameters: $M=2t=2\xi=1$ and $\Delta=$ 0 and 0.25 for the TI and HOTI phases, respectively. 
In Fig. \ref{HOTIxTI} we compare the electronic structure of 2D, 1D and 0D structures in the TI and HOTI phases. The band structure for the (a) TI and (e) HOTI is shown along with the spin Hall conductivity (SHC). 
In both cases, a 2 eV band gap is obtained and the band structures are very similar, with an exception along the M$-$Y path. However, as the TI shows a quantized ($e^2/h$) SHC, the HOTI differs with a value $\approx$ 0.75 $e^2/h$. The HOTI SHC is gradually reduced for higher $\Delta$ values, as presented in the main text. 
The panels (c) and (g) shows the 0D (square nanoflake) structure electronic eigenvalues for the TI and HOTI phases, respectively. 
For the TI, countless midgap states are obtained whereas in the HOTI, only eight perfectly degenerate states are obtained. 
The spatial localization of these midgap states are remarkably different. Fig. \ref{HOTIxTI} (d) and (h) shows the midgap states wave function real space projection, for the TI and HOTI, respectively. The TI midgap states are located all along the nanoflake edge (yellow and green regions). Conversely, the HOTI midgap states are mainly located at the nanoflake corners. The corner states localization will be discussed in the next section.

\clearpage

\subsection{Corner States Localization}
\begin{figure*}[ht!]
\includegraphics[width=\textwidth]{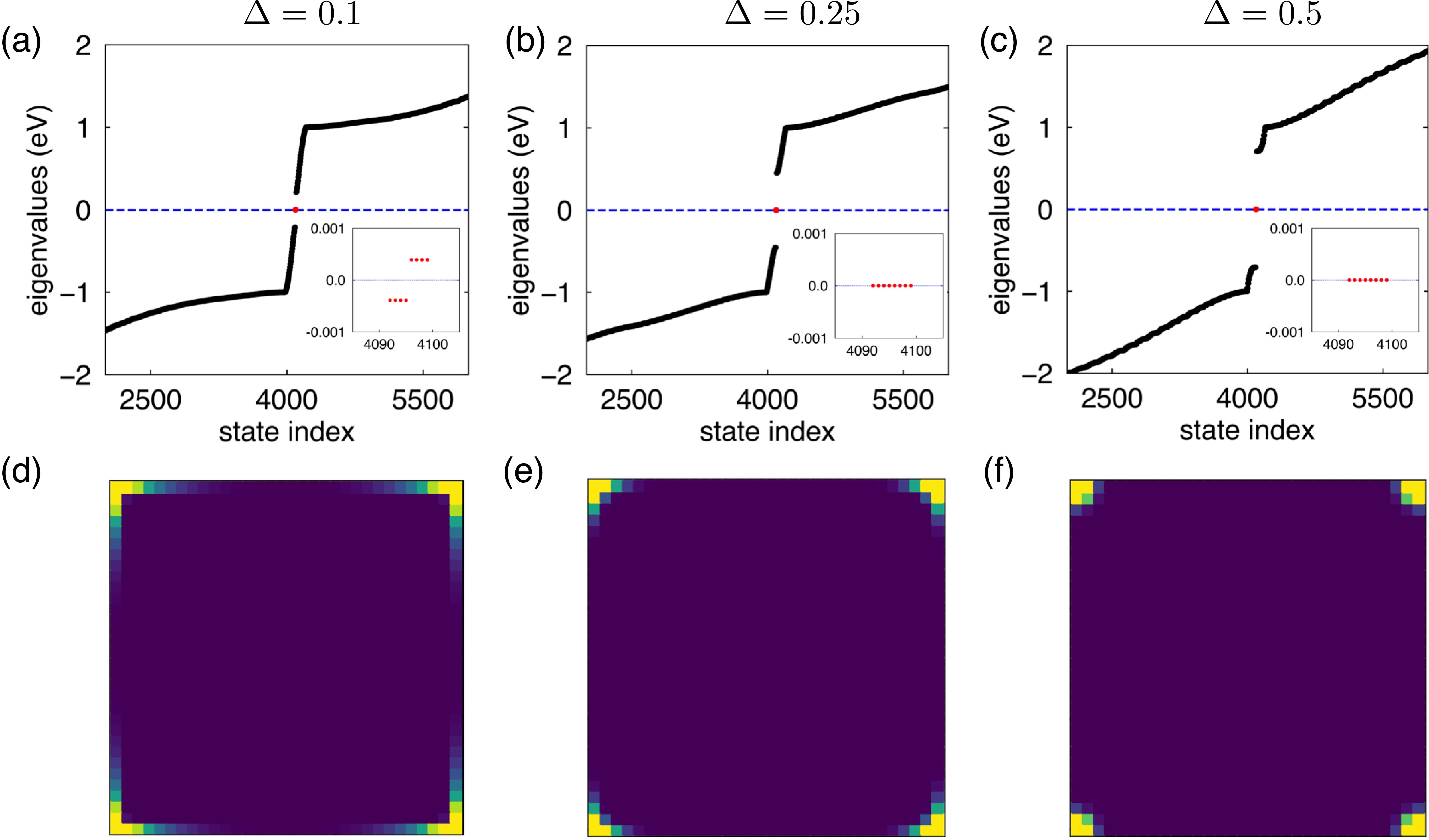}
\caption{\label{projections-delta} 2D HOTI square nanoflake $\Delta$ effect. Tight-binding model parameters are $M=2t=2\xi=1$. (a-c) Eigenvalues for $\Delta=$ 0.1, 0.25 and 0.5, respectively. The inset shows a narrow energy window around the corner states (in red). (d-f) Corner states wave function real space projection for $\Delta=$ 0.1, 0.25 and 0.5, respectively. The nanoflake is composed of 32$\times$32 lattice sites.}
\end{figure*}

Here we will use the HOTI tight-binding model to understand why the triangular nanoflakes corner states are not degenerate. In Figure~\ref{projections-delta} we show the 2D HOTI tight-binding model square nanoflake eigenvalues. The model parameters are $M=2t=2\xi=1$, and different $\Delta$ values of (a) 0.1, (b) 0.25 and (c) 0.5 are used. 

In the model, nonzero $\Delta$ values implies a TI $\rightarrow$ HOTI topological phase transition. Here we define a quantity called corner states spread ($\gamma$) as the energy difference between its the highest and lowest eigenvalue. As $\Delta$ increases, $\gamma$ becomes small until fully degenerate at the Fermi level, see the Figure insets. The interpretation of the above results is straightforward: for materials close to the TI-HOTI phase transition, {\it i.e.} small $\Delta$, the corner states are less localized, implying in degeneracy breaking. This can be corroborated by inspecting the corner states wave function real space projection shown in Figure~\ref{projections-delta}(d-f), where $\Delta$ increases like given in (a-c). The color code is: larger (smaller) values are indicated by the yellow (purple) colors. The increased localization is clear evaluating panels (d-f).

\begin{figure*}[ht!]
\includegraphics[width=\textwidth]{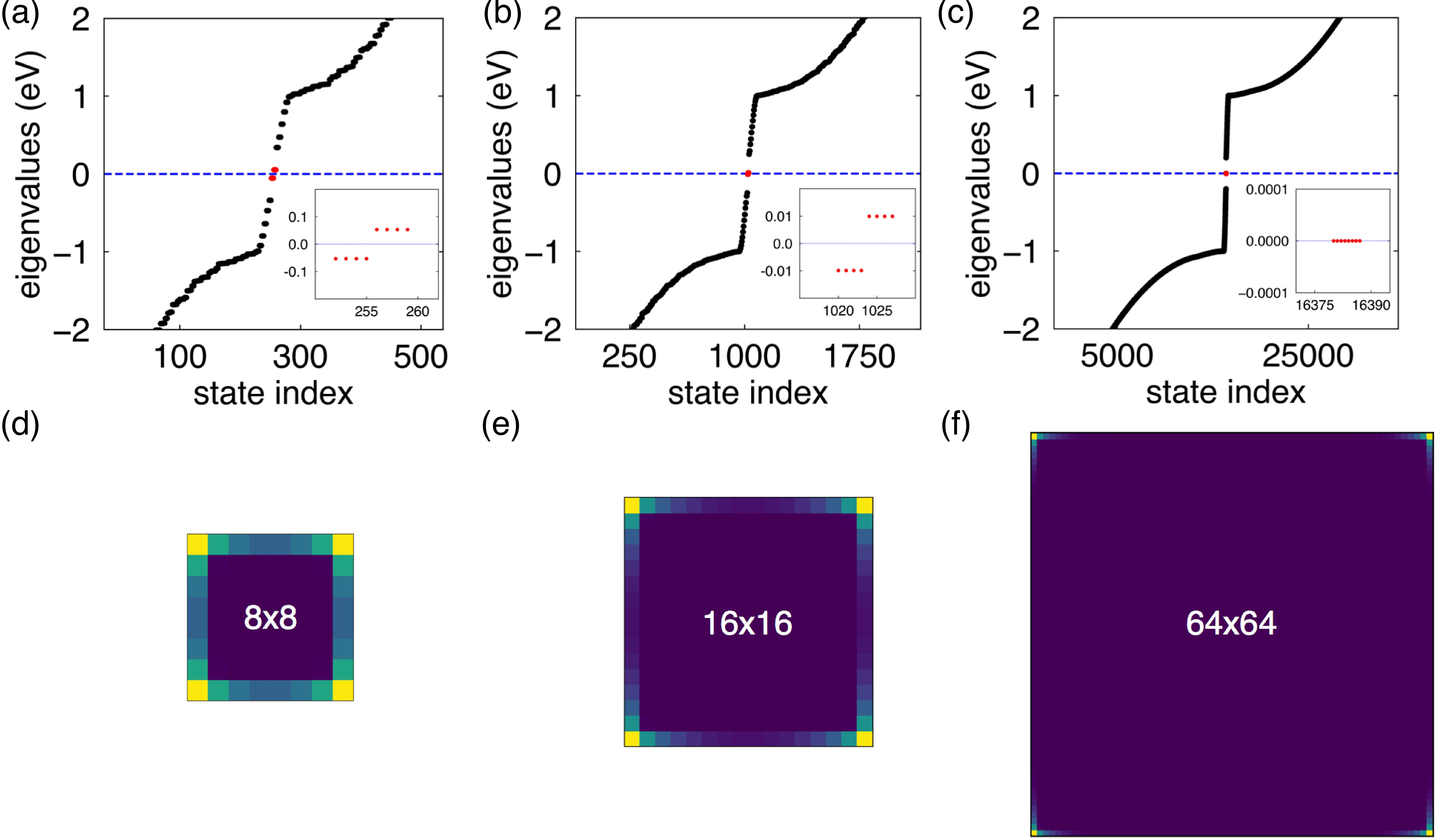}
\caption{\label{projections-lattice} 2D HOTI square nanoflake size effect. Tight-binding model parameters are $M=2t=2\xi=1$ and $\Delta=$ 0.1. (a-c) Eingenvalues for a nanoflake with 8$\times$8, 16$\times$16, and 64$\times$64 lattice sites, respectively. The 32$\times$32 nanoflake is also shown in Figure~\ref{projections-delta}(a). The inset shows a narrow energy window (different scales) near the corner states (in red). (d-f) Corner states wave function real space projection for each corresponding nanoflake. Larger (small) projection values are indicated by the yellow (purple) color.}
\end{figure*}

To deepen our analysis, the $\Delta$ parameter is kept fixed at 0.1 and the nanoflake area is increased until 64$\times$64 lattice sites. In Figure \ref{projections-lattice}(a-c) the HOTI model eigenvalues are shown for 8$\times$8, 16$\times$16, and 64$\times$64 lattice sites, respectively. The 32$\times$32 case was already shown in Figure \ref{projections-delta}(a). The 8$\times$8 nanoflake spread $\gamma$ defined above is around 0.2 eV, and as the area increases the spread is reduced until the corner states become fully degenerate at the Fermi level. Nevertheless, differently from increasing $\Delta$, the corner states localization is unchanged. Therefore, the larger the distance between the nanoflake vertices, the lesser is their interaction. The main consequence of using nanoflakes with reduced sizes is the corner states broken degeneracy.

A common interpretation is that the 2D topological corner states are critically localized and can not participate in transport phenomena. Here we demonstrate that when SOC is strong this notion is not correct. This delocalization of corner states was observed before, Ezawa has demonstrated nonzero midgap electrical conductance in a TB model in the presence of SOC~\cite{Ezawa2018a}. Where, starting from a TI, the HOTI phase is induced via an in-plane magnetic field, which would be analogous to the mixing term $\Delta$ in our model~\cite{Ezawa2018a}. 

\begin{figure*}[h!]
\includegraphics[width=\textwidth]{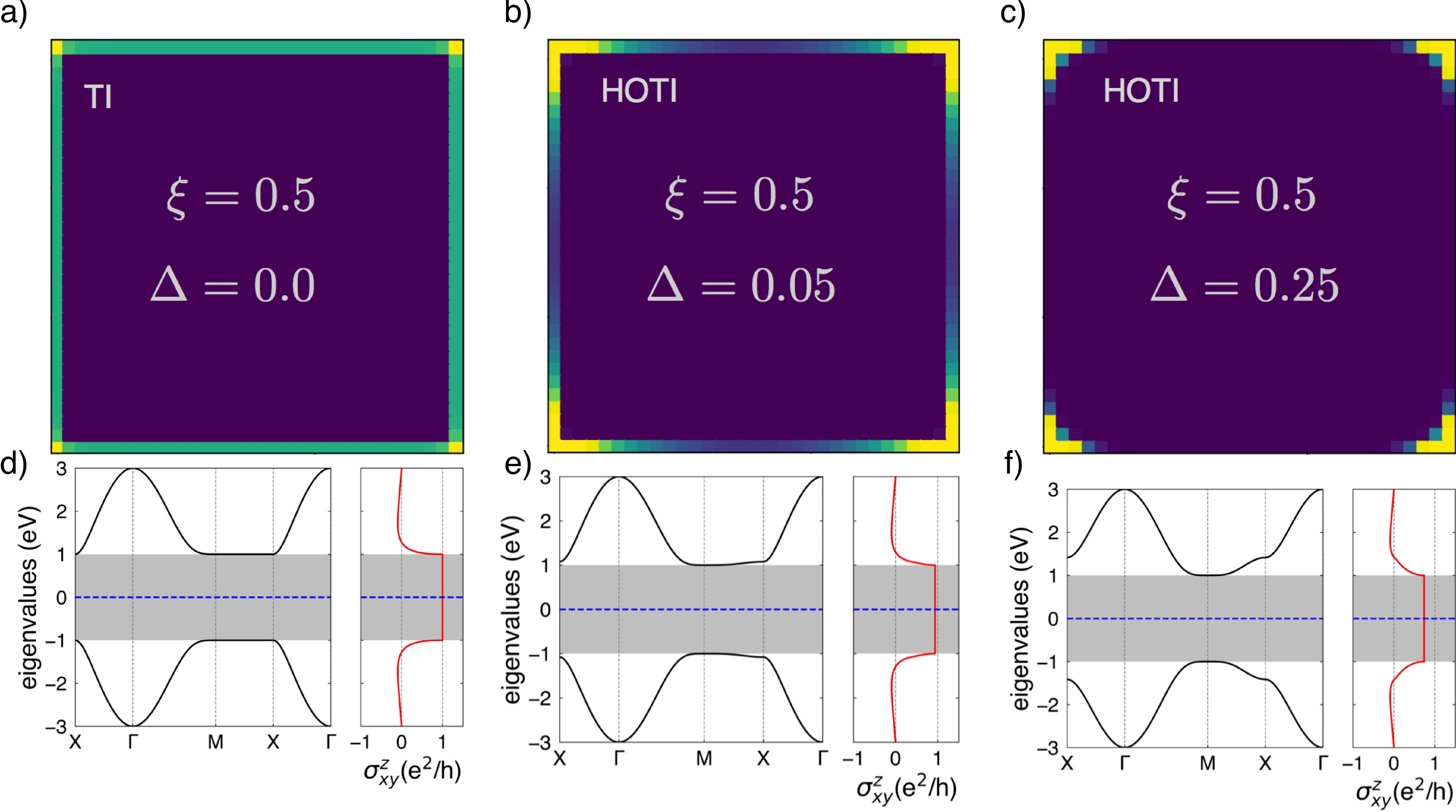}
\caption{\label{deltaSHC} 2D HOTI square nanoflake mixing parameter $\Delta$ effect: (a,d) $\Delta=0.0$ corresponding to a TI; (b,e) $\Delta=0.05$ corresponding to a HOTI; (a,d) $\Delta=0.25$ corresponding to a HOTI. Tight-binding model parameters are $M=2t=1$, and $\xi=0.5$. (a-c) Corner states wave function real space projection for each corresponding nanoflake. Larger (small) projection values are indicated by the yellow (purple) color. (d-e) Bulk band structure along with the spin Hall conductivity (SHC).}
\end{figure*}

Exploring the model parameters, spin-orbit coupling strength ($\xi$) and mixing term ($\Delta$), we can control the corner states localization and thus increasing (decreasing) the midgap SHC.
Fig. \ref{deltaSHC}(a-c) shows the corner states localization for different $\Delta$ values, indicating the HOTI degree ($\Delta=0$ equals to a TI):  a) 0.05, b) 0.15 and c) 0.25. The SOC parameter is kept fixed at 0.5. 
We compute the midgap states localization for the TI phase, i.e. $\Delta=0$, which is also shown in Fig. \ref{HOTIxTI}(a-d).
For the TI the topological states are localized at the nanoribbon edges, as expected.
In the lower panel, Fig. \ref{deltaSHC}(d-f), is the corresponding bulk bandstructure and SHC. For $\Delta=0.05$, the corner states are still very delocalized, resulting in a large bulk SHC. As $\Delta$ increases, the corner states become increasingly localized at the nanoflake corners, and its bulk SHC is reduced. Therefore, we establish that the SHC is inversely proportional to the $\Delta$ parameter, which for real materials is a measure of the HOTI degree.

\begin{figure*}[h!]
\includegraphics[scale=0.06]{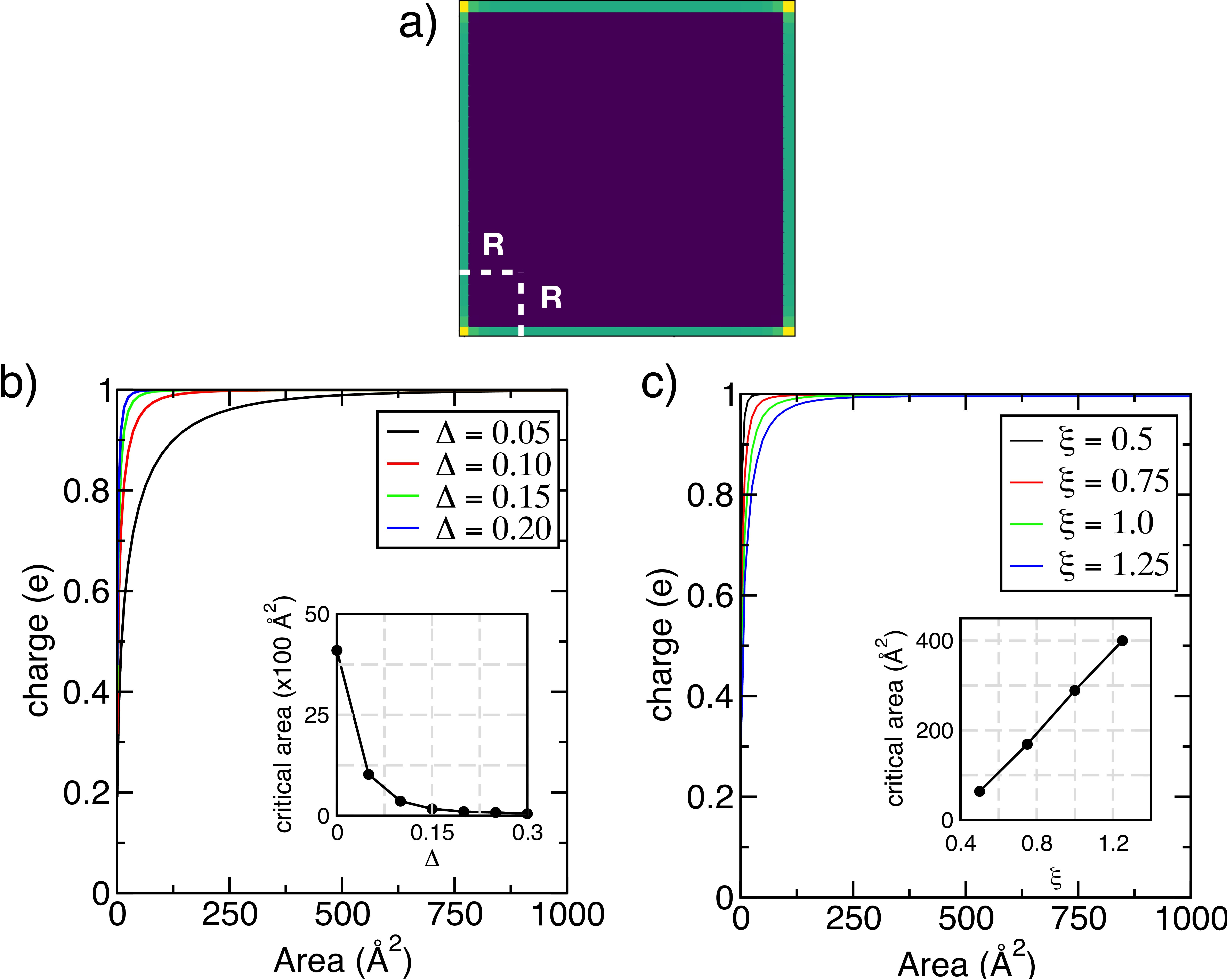}
\caption{\label{critical-area} 2D HOTI square nanoflake charge integration. Tight-binding model parameters are $M=2t=1$. (a) Definition of integration area (R$^2$). Charge as a function of integrated area for (b) $\Delta=$ 0.05, 0.10, 0.15, 0.20 and (c) $\xi=$ 0.5,0.75, 1.0, 1.25. The critical area is defined as the area where the integrated charge is equal to one.}
\end{figure*}
In Fig.~\ref{critical-area} we calculate the corner states delocalization by integrating its charge. Starting from a selected given corner we increase the integrated area, panel (a), until the integrated charge is equal to one, which is defined as a critical area. In panel (b) we fixed the SOC parameter at 0.5 and vary the mixing term from 0.05 to 0.20. What we observe is that the corner states become ever more localized, which will reflect in the reduction of the SHC. On the contrary, as the SOC parameter becomes larger, for a fixed $\Delta$ of 0.25, the corner states are less localized and thus the SHC is larger, panel (c). The critical area for the TI phase, i.e. $\Delta=0.0$, is the whole nanoflake area, since TIs topological states are fully delocalized along its edges.

\clearpage

\subsection{DFT $\times$ \textsc{paoflow} Bandstructure}
\begin{figure*}[h!]
\includegraphics[scale=0.08]{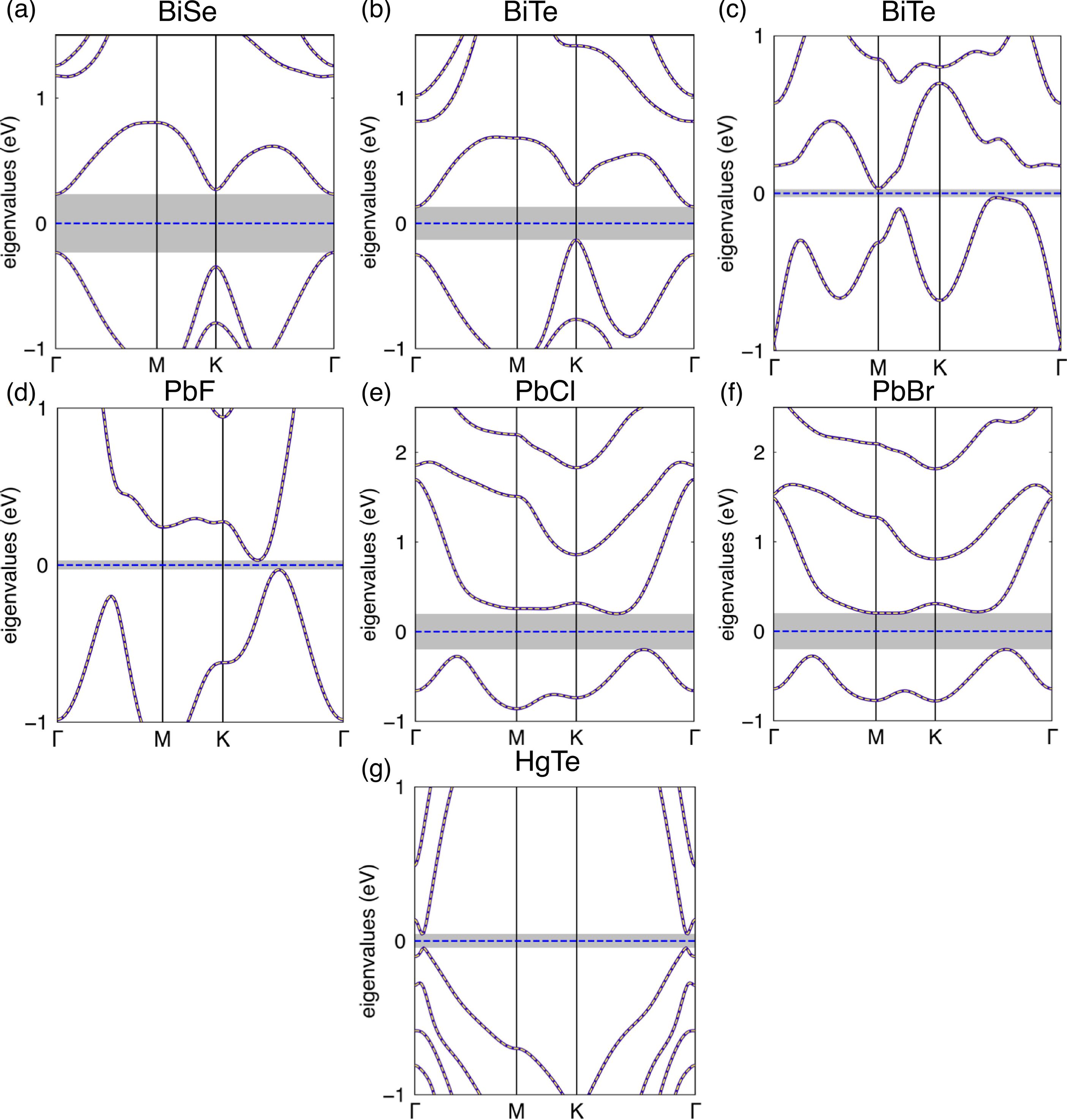}
\caption{\label{DFTxPAO} 2D HOTI band structure : DFT $\times$ \textsc{PAOflow}. (a) BiSe and (b) BiTe in the GaSe prototype structure. (c-g) BiTe, PbF, PbCl, PbBr and HgTe in the CH prototype structure, respectively. The solid purple (dashed yellow) line represents the DFT (\textsc{paoflow}) band structure.}
\end{figure*}

In Figure \ref{DFTxPAO}(a-g) the predicted HOTI materials DFT and \textsc{paoflow} fully relativistic band structures are compared. The DFT (solid purple line) and \textsc{paoflow} (dashed yellow line) are in excellent agreement. This methodology is very efficient since it reduces the 
basis functions subspace from several thousand of plane waves to a few atomic orbitals functions. As an example, the BiSe (GaSe prototype) is composed of only 72 atomic orbitals functions, 18 orbitals per Bi/Se atom.

\clearpage

\subsection{BiTe on CH Prototype Band Structure}
\begin{figure*}[ht!]
\includegraphics[width=\textwidth]{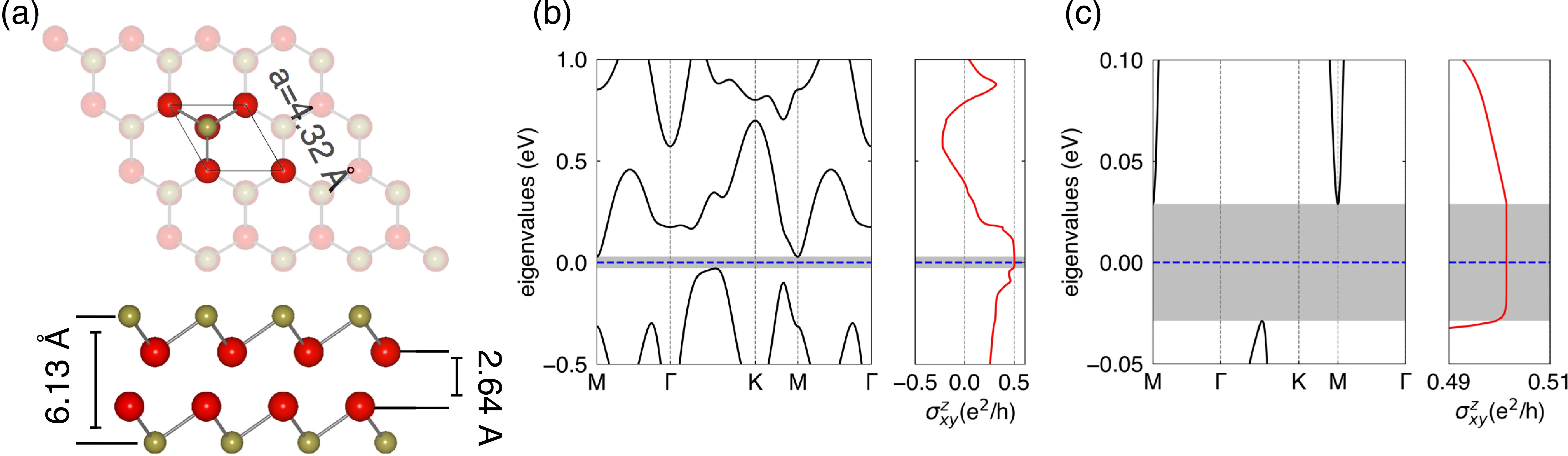}
\caption{\label{BiTe-CH} Two-dimensional HOTI material BiTe (in the CH prototype). (a) 2D bulk crystal structure top and side views. (b) Bulk band-structure along the high symmetry lines and the spin Hall conductivity ($\sigma^{z}_{xy}$) as a function of the energy. (c) Same as (b) with a narrower energy window. The dashed blue line represents the Fermi level. The gray shaded area encompass the bulk band gap. Due to the BiTe small bandgap we increased the $K$-point mesh for the spin Hall conductivity calculation by a factor of two. The nanoflake geometry eigenvalues is not shown because the small bandgap would require an extremely large supercell.}
\end{figure*}

\clearpage

\subsection{PbF on CH Prototype Band Structure}
\begin{figure*}[ht!]
\includegraphics[width=\textwidth]{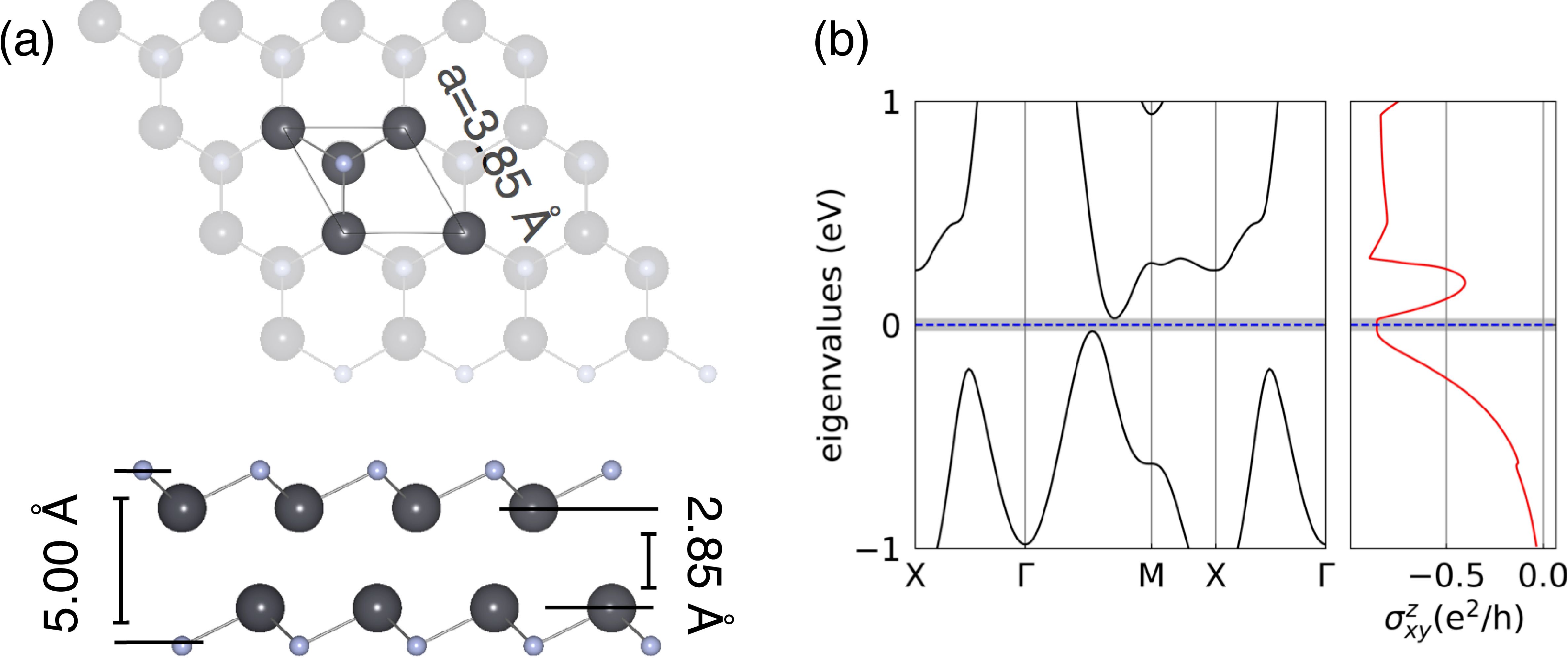}
\caption{\label{PbF-CH} Two-dimensional HOTI material PbF (in the CH prototype). (a) 2D bulk crystal structure top and side views. (b) Bulk band-structure along the high symmetry lines and the spin Hall conductivity ($\sigma^{z}_{xy}$) as a function of the energy. The dashed blue line represents the Fermi level. The gray shaded area encompass the bulk band gap. Due to the PbF small bandgap we increased the $K$-point mesh for the spin Hall conductivity calculation by a factor of two. The nanoflake geometry eigenvalues is not shown because the small bandgap would require an extremely large supercell.}
\end{figure*}

\clearpage

\subsection{HgTe on CH Prototype Band Structure}
\begin{figure*}[ht!]
\includegraphics[width=\textwidth]{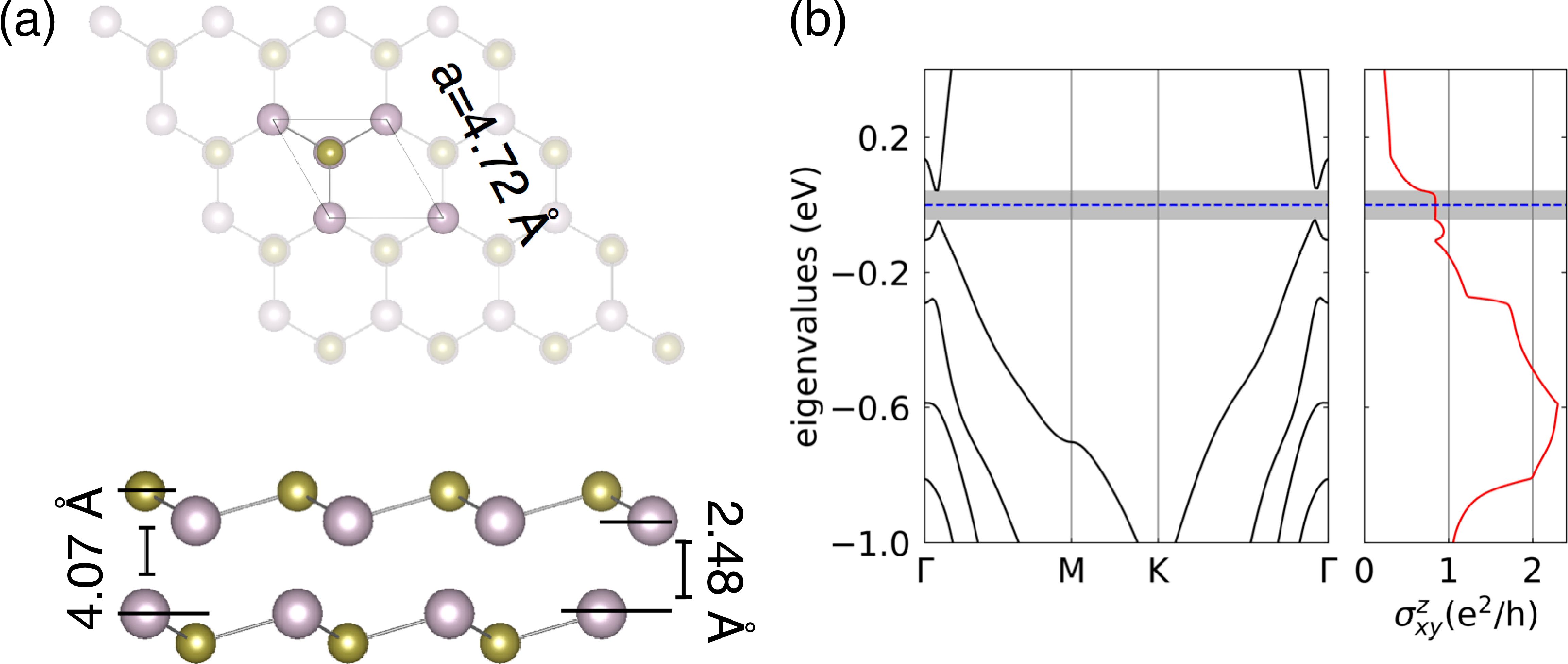}
\caption{\label{HgTe-CH} Two-dimensional HOTI material PbF (in the CH prototype). (a) 2D bulk crystal structure top and side views. (b) Bulk band-structure along the high symmetry lines and the spin Hall conductivity ($\sigma^{z}_{xy}$) as a function of the energy. The dashed blue line represents the Fermi level. The gray shaded area encompass the bulk band gap. Due to the HgTe small bandgap we increased the $K$-point mesh for the spin Hall conductivity calculation by a factor of two. The nanoflake geometry eigenvalues is not shown because the small bandgap would require an extremely large supercell.}
\end{figure*}

\clearpage

\subsection{BiTe on GaSe Prototype Band Structure}
\begin{figure*}[ht!]
\includegraphics[width=\textwidth]{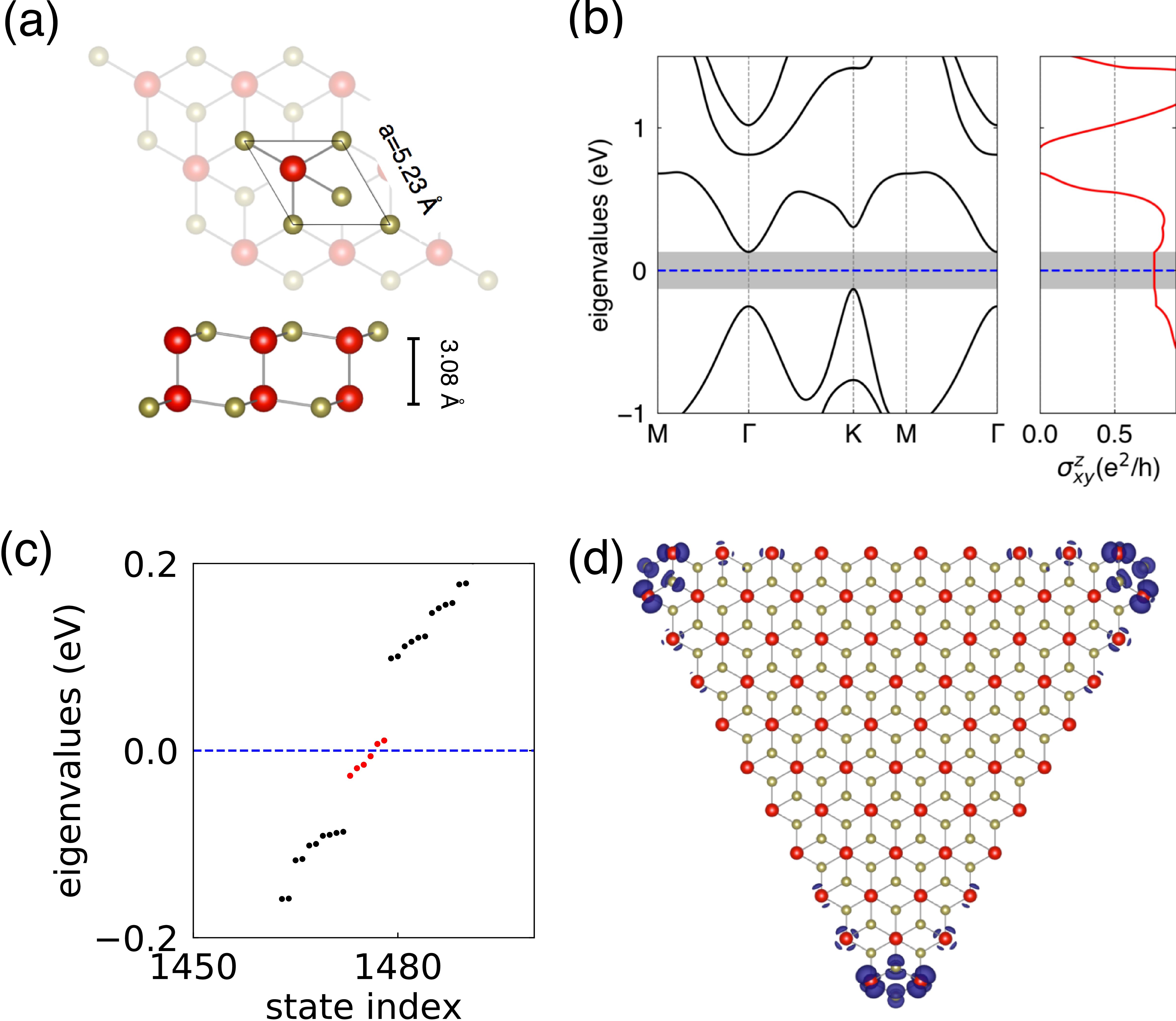}
\caption{\label{BiTe-GaSe} BiTe in the GaSe prototype. ({a}) crystalline structure, ({b}) bulk band structure and spin Hall conductivity, ({d}) flake structure with corner states charge density, and ({c}) corresponding 0D eigenvalues.}
\end{figure*}

\clearpage
\subsection{PbCl on CH Prototype Band Structure}
\begin{figure*}[ht!]
\includegraphics[width=\textwidth]{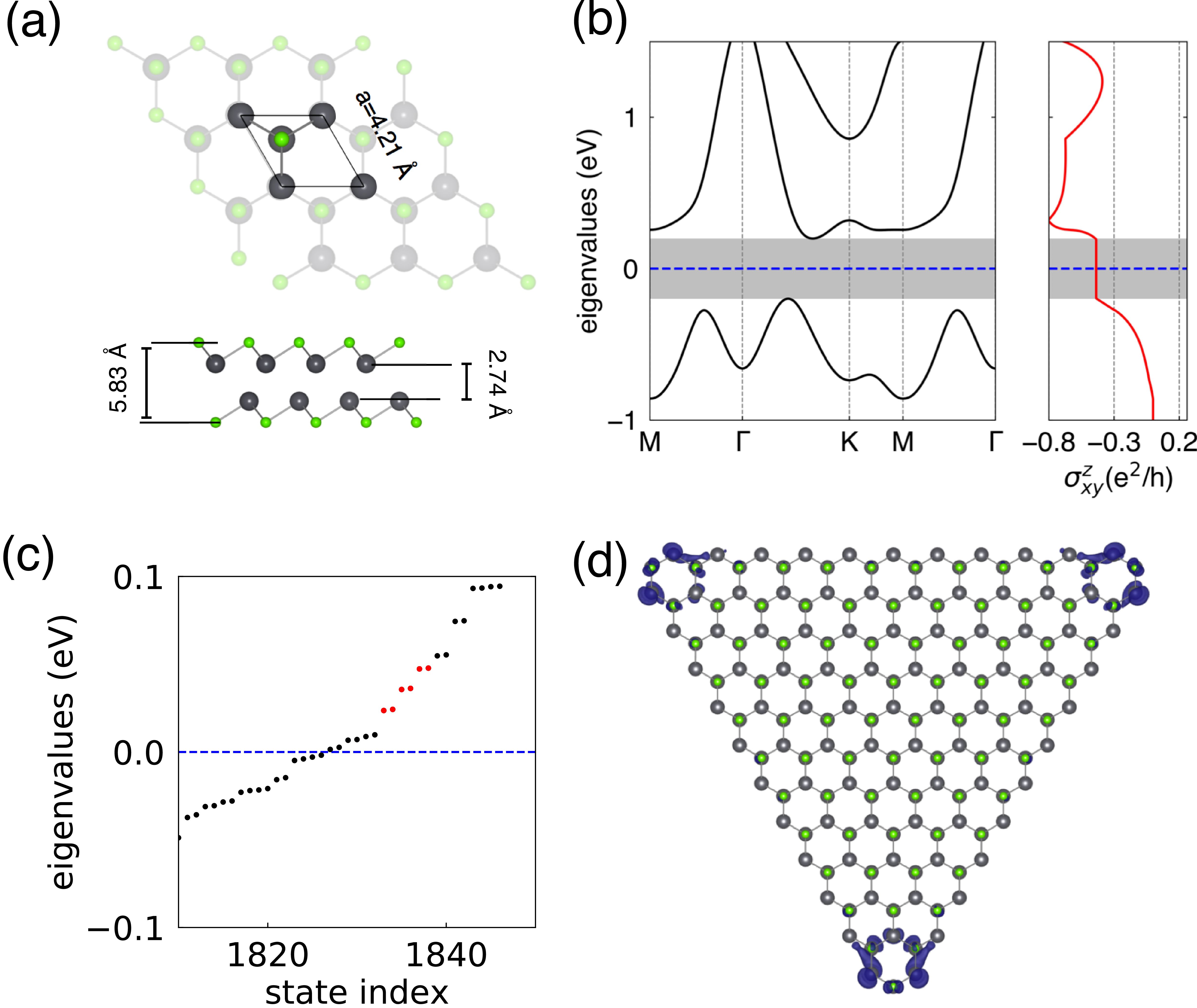}
\caption{\label{PbCl-CH} PbCl in the CH prototype. ({a}) crystalline structure, ({b}) bulk band structure and spin Hall conductivity, ({d}) flake structure with corner states charge density, and ({c}) corresponding 0D eigenvalues.}
\end{figure*}

\normalbaselines
%